\documentclass[11pt]{article}
\usepackage{amsmath,epsfig,url,centernot, graphicx}
\usepackage{subcaption}
\usepackage{xcolor}
\begin{document}
\baselineskip=17.5pt
%\baselineskip=15pt

% function for conditional independence - from link above
\newcommand{\bigCI}{\mathrel{\text{\scalebox{1.07}{$\perp\mkern-10mu\perp$}}}}
\newcommand{\nbigCI}{\centernot{\bigCI}}

\newcommand{\CI}{\mathrel{\perp\mspace{-10mu}\perp}}
\newcommand{\nCI}{\centernot{\CI}}

\newcommand{\sn}{\sum_{i=1}^n}
\newcommand{\snj}{\sum_{j=1}^n}
\newcommand{\intt}{\int_0^t}
\newcommand{\inti}{\int_0^{\infty}}
\newcommand{\inttau}{\int_0^{\tau}}

\begin{center}
{\Large \bf  Bridging linearity-based and kernel-based sufficient dimension reduction} \\
\end{center}

\vspace{.001mm}
\begin{center}
Youngjoo Cho$^1$ and Debashis Ghosh$^2$\\
$^1$ Department of Mathematical Sciences \\
The University of Texas at El Paso, El Paso, TX  79968\\
ycho@utep.edu\\
$^2$Department of Biostatistics and Informatics\\
Colorado School of Public Health, Aurora, CO, 80045\\
debashis.ghosh@cuanschutz.edu\\
\end{center}

\vspace{.001mm}

\begin{center}
{\large \bf \ Summary}
\end{center}

There has been a lot of interest in sufficient dimension reduction (SDR) methodologies as well as nonlinear extensions in the statistics literature.   
%The SDR methodology has been motivated by several considerations: (a) finding data-driven subspaces that capture the essential facets of regression relationships; (b) analyzing data in a `model-free' manner.   
In this note, we use classical results regarding metric spaces and positive definite functions to link linear SDR procedures to their nonlinear counterparts.  
\\
\\
{\bf Keywords:} Central subspace; Kernel Methods; Machine Learning; Semiparametric model; Sliced Inverse Regression.
\newpage

\section{Introduction}

\noindent 
In statistical modelling, a key challenge is to determine appropriate transformations of the data that can reduce its dimension while at the same time capturing the essential information in the regression relationship between a set of covariates and a response variable.   Towards this end, there has been a field of statistics, termed sufficient dimension reduction (SDR), that has sought to develop methodology with this goal in mind.    Broadly speaking, sufficient dimension reduction represents a class of `model-free' methodologies that seek to find directions in the data that can capture the essential information in the regression relationship previously mentioned.  

Historically, the basis for sufficient dimension reduction methods was the observation by authors such as Brillinger (1977) and Li and Duan (1989) that regresion parameters estimated by ordinary least squares were consistent, up to a constant, for regression parameters in a generalized single-index model.  This result required an assumption on the covariates being elliptically symmetric, which has been reframed into the current sufficient dimension reduction literature as the so-called linearity assumption.   More recent formulations for sufficient dimension reduction have postulated the existence of a central subspace; subsequently the goal of sufficient dimension reduction methods is to estimate the basis vectors of the central subspace.  Many SDR methods (Li, 1991; Li, 1992, Cook and Ni, 2005) fall into this category.  
%such as sliced inversion regression SIR) algorithm proposed by Li 1991), Principal Hessians Directions Li, 1992), an optimization of an objective function developed by Cook and Ni 200?), minimum adaptive variance estimation Xia et al., 2002) and a semiparametric estimation procedure recently developed by Ma and Zhu 2012).   

As mentioned above, one of the key assumptions in applying the sufficient dimension reduction methodology in the previous paragraph is termed the linearity condition.    A sufficient condition for this to hold is that the predictor variables of interest follow an elliptically contoured distribution.  Distributions that satisfy elliptical symmetry include the multivariate normal distribution and the multivariate t-distribution.  We term any SDR method that requires the linearity condition linearity-based SDR.   One of the main criticisms levelled against linearity-based SDR methods is that this assumption will not be satisfied in practice.  For example, if covariates are discrete, then this will violate the linearity condition.   Many authors invoke the theoretical results of Hall and Li (1993), which suggest that in an asymptotic framework, the linearity condition will hold.  

An alternative for SDR has been kernel-based SDR approaches.  Such approaches can be found in Ferr\'e and Yao (2003), Fukumizu et al. (2004, 2009), Wu (2008),  Wu et al. (2013), Li et al. (2011) and Lee et al. (2013).  The approaches of Li et al. (2011) and Lee et al. (2013) completely avoid the linearity condition by formulating a more general notion of sufficiency and develop estimation procedures based on constructional of kernel matrices and eigenvalue/eigenvector decomposition to estimate fitted functions.

What we propose in this article is a link  between linearity-based and kernel-based SDR methodology.   To do this, we use the theory of positive definite functions that was initially developed by Schoenberg (1938).  This allows for a framework in which the dimension of the predictor variables tends to infinity, which is very appealing to current scientific fields such as genetics and neuroimaging.   This allows for a characterization using a specific class of kernels.   %This equivalence allows for estimation of directions using a penalized estimation framework and allows for generalization to other loss functions and functionals of interest.  Such a framework also gives a new perspective as to what the limiting functional of sufficient dimension reduction procedures will be. 
This result, summarized as a Proposition in Section 3, shows that the linearity condition is not as problematic as it appears on first glance.

\section{Data Structures and Review of Sufficient Dimension Reduction Methods}

Let the data be represented as $(Y_i,{Z}_i)$, $i=1,\ldots,n$, a
random sample from the triple $(Y,{Z})$, where $Y$
denotes the response of interest and $Z$ is a $p$-dimensional vector of covariates.  
Suppose we formulate the following regression model for $Y$ given ${Z}$: 
\begin{equation}\label{semiprop}
 E(Y\mid{Z}) = g(\beta_1'{Z},\beta_2'{Z},\ldots,\beta_k'{Z}, u),
 \end{equation}
where $\beta_j$ $(j=1,\ldots,k)$ are $p-$dimensional vectors of unknown regression coefficients, $u$ is an
error term, and $g$ is an unspecified monotonic link function.  Because of the presence of the parametric components involving $\beta_j$ as well as the nonparametric specification of the link function, model (\ref{semiprop}) is semiparametric.   Note that when $k = 1$, model (\ref{semiprop}) reduces to a single-index model.  In addition, the model (\ref{semiprop}) can accommodate non-homoskedasticity in the error term if the variance depends on $\beta_j'{Z}$.

The starting point of sufficient dimension reduction methods is the conditional
independence of $Y$ and ${Z}$ given $E(Y\mid{Z})$.  An implication of 
model (\ref{semiprop}) being true is that
there exists a $p \times k$ matrix {B}, where
\begin{equation}\label{ci1}
Y \bigCI Z \mid {B}'{Z}
\end{equation}
Another way of stating (\ref{ci1}) is that the projection ${B}'{Z}$ provides a sufficient data reduction and contains the
essential information about the relationship between $Y$ and ${Z}$.
More generally, we can define a projection operator ${P}_B$ to
be the projection operator onto the subspace spanned by the
columns of ${B}$.  Then (\ref{ci1}) can be reexpressed as
\begin{equation}\label{ci2}
Y \bigCI Z \mid {P}_{B}{Z}.
\end{equation}
If (\ref{ci2}) holds, then it also holds for any subspace ${C}$ such that the span of ${B}$ is the same as the span of
${C}$.  Let $S({B})$ be the subspace generated by the
columns of ${B}$.   Let $S_{*}$ denote the
intersection of all possible subspaces; if
$S_{*}$ is also a subspace, i.e. it satisfies
(\ref{ci2}), then we will refer to $S_{*}$ as the
central subspace (Li, 1991;
Cook, 1998).  We will assume throughout that the central 
subspace exists (Cook, 1998; Yin, Li and Cook, 2008).   As mentioned in the Introduction, there are many algorithms available for estimating the basis vectors of the central subspace using linearity-based SDR (Li, 1991; Li, 1992, Cook and Ni, 2005).

An alternative to linearity-based SDR is kernel-based SDR (Ferr\'e and Yao, 2003; Wu, 2008; Wu et al., 2013; Fukumizu et al., 2004, 2007, 2009; Li et al., 2011; Lee et al., 2013).   Most of these proposals use reproducing kernel Hilbert spaces (Wahba, 1990; Berlinet and Thomas-Agnan, 2004) in order to construct generalizations of basis vectors for the central subspace. 
%This generalization was suggested in remark 2.4. of Li (1992). 
%This class of functions will have an associated kernel structure that is given in Theorem 1.  One major difference is that our methodology constrains the classes of kernels that can be used,  while the previous proposals required a different type of property termed characteristic kernel Srirempubudur et al., 2011).   

\section{Linking sufficient dimension reduction procedures}

While the starting points of linearity-based and kernel-based SDR methods appear different at first glance, we now show how the two are connected in this section.  To prove the main result in this section, we will require the definitions of positive definite and completely monotone functions.   
\\
\noindent {\bf Definition 1.}  A real-valued function $f$ is said to be positive definite if for any set of real numbers $x_1,\ldots,x_n$, the $n \times n$ matrix A with $(i,j)$th entry $a_{ij} = f(x_i - x_j)$ $(i=1,\ldots,n; j=1,\ldots,n)$ is positive definite.   
\\
\noindent {\bf Definition 2.} A real-valued function $f$ is said to be completely monotone if for all $r \in \{0,1,2,\ldots\}$, 
$$ (-1)^r f^{(r)}(x) \geq 0,$$
where $f^{(r)}$ denotes the $r-$th derivative of $f$.   

A function $f(t)$ $(t \in R)$ is positive definite if and only if $f(t) = g(t^2)$, where $g$ is completely monotone.  The other key fact is that any positive definite function will define a kernel (Aronszajn, 1950).  Thus, for any positive definite function  $f$, we have that $K({Z},\tilde{Z}) = f(\|{Z}  - \tilde{Z}\|)$ is a proper kernel.   

% To do this
%will require the notion of a positive definite function.   Let ${\cal F}$ denote a Hilbert space with an associated norm $\\mid \ \ \\mid_{\cal F}$.   
%A example of a finite-dimensional HIlbert space is $R^p$ with the norm being defined based on Euclidean distance.   
%A positive definite function is a function $f: {\cal F} \rightarrow R$ such that for any $m$ vectors ${A}_1,\ldots,{A}_n \in {\cal F}$, the matrix ${A}$ with $i,j)$th element $A_{ij} = [   f\\mid{A}_i - {A}_j\\mid^2_{\cal F})]$ is positive definite.   
As in Schoenberg (1938), we will study spaces of positive definite functions that are defined on proper metric spaces.   The space $R^p$ with the Euclidean distance can also be viewed as a metric space.  Let ${\cal B}(E)$ denote the space of positive definite functions for a metric space $E$.   One result of Schoenberg (1938) was that if $E_1$ and $E_2$ are metric spaces with $E_1 \subset E_2$, then ${\cal B}(E_1) \supset {\cal B}(E_2)$.   If we take $E_2$ to be the restriction of $R^p$ to random vectors ${Z}$ that satisfy the linearity condition and $E_1$ to be random vectors which are elliptically symmetric, then we have ${\cal B}(E_1) \supset {\cal B}(E_2)$.  For ${\cal B}(E_2)$, we have the following characterization from Schoenberg 1938): 

\noindent {\bf Lemma 1.}   A $p-$dimensional random vector ${W}$ is elliptically symmetric if and only if its characteristic function can be written as $\psi(\|{ w}\|^2)$, where ${w} \in R^p$ and $\psi(t)$ has the form
\begin{equation}\label{es}
 \psi(t) = \int_0^{\infty} \omega_p (r^2t) dF(r),
 \end{equation}
where $\omega_p$ is the characteristic function for a $p-$dimensional random vector that is distributed uniformly on the unit sphere in $R^p$, and $F(r)$ is a distribution function on $[0,\infty)$.  We note that the form of $\omega_p(t)$ is given by
$$ \omega_p(t) = \Gamma \left ( \frac{p}{2} \right ) \left  (\frac{2}{t} \right )^{p-2)/2} J_{(p-2)/2}(t),$$
where $\Gamma(a) \equiv \int_0^{\infty} u^{a-1}\exp(-u) du $ denotes the Gamma function and 
$$J_{\alpha}(x) \equiv \sum_{m=0}^{\infty} \frac{(-1)^m}{m!\Gamma(m+\alpha+1)} \left ( \frac{x}{2} \right )^{2m+\alpha}$$
represents the Bessel function.   

Given the definitions of $E_1$ and $E_2$ above, we define a sequence of metric spaces in the following way: let $E_{2+i}$ be a metric space consisting of elliptically symmetric random vectors in $R^{p+i}$ for $i=1,2,\ldots,$.   We have that elliptical symmetry in higher dimensions implies elliptical symmetry in lower dimensions.  This yields the following chain of inclusion relations:
\begin{equation}\label{inequalities}
{\cal B}(E_1) \supset {\cal B}(E_2) \supset {\cal B}(E_3) \supset \cdots \supset  {\cal B}(E_{\infty}).
\end{equation}
In addition, Schoenberg (1938) provides a characterization of ${\cal B}(E_{\infty})$ in (\ref{inequalities}), which is given in the following result:

\noindent {\bf Lemma 2.} A random element ${\cal W}$ exists in ${\cal B}(E_{\infty})$  if and only if its characteristic function can be written as 
$\psi(\|{w}\|^2)$, where $\psi(t)$ has the form
\begin{equation}\label{eshilbert}
 \psi(t) = \int_0^{\infty} \exp(-r^2t) dF(r), \ \ t >0
 \end{equation}
and $F(r)$ is a distribution function on $[0,\infty)$.  

\noindent {\bf Remark 1.}   Note that by the nested structure of the space of positive definite functions in (\ref{inequalities}), it is also the case that
$${\cal B}(E_{\infty}) = \cap_{i=1}^{\infty} {\cal B}(E_i).$$
Thus, ${\cal B}(E_{\infty})$ is the smallest space containing ${\cal B}(E_i)$ for all $i$.  In this sense, the object ${\cal B}(E_{\infty})$ can be interpreted as an infinite-dimensional analog to the central subspace that was described in \S 2.   Combining all the results above leads us to the following result.

\noindent {\bf Proposition.}   A random element exists in ${\cal B}(E_{\infty})$ if and only if its associated kernel is of the form 
\begin{equation}\label{tinv}
K({X},\tilde {X}) = \psi(\|{X} - \tilde{X}\|),
\end{equation}
where $\psi$ is generated via (\ref{eshilbert}).   The proposition shows that the kernels in ${\cal B}(E_{\infty})$ only depend on the interpoint distances between points.  

\noindent {\bf Remark 2.} 
We note that we arrive at kernels as in Lee et al. (2013) but with a very different starting point and a different set of assumptions.   We do so through an alternative construction that did not rely on the generalized notions of sufficiency that are considered by Lee et al. (2013). The nesting function space argument in this paper allows one to transition from distributional assumptions (e.g., elliptical distribution for $X$) to functional definitions that can be characterized using kernels.   

\noindent {\bf Remark 3.}   We recall an earlier example from the SDR literature that violates the linearity condition.  The example is the regression model 
$$ Y = (\beta'X)^2 + \epsilon,$$ where $X$ and $\epsilon$ have normal distributions.  A method such as SIR will estimate the direction to be zero.   Using the theoretical framework that is presented here, we would see that this regression relationship would not exist in ${\cal B}(E_{\infty})$.   Formally, the regression model would correspond to a kernel of the form $K(X,\tilde X) = (<X,\tilde X>+1)$ which has been referred to as the polynomial kernel of order one in the machine learning literature.   By the proposition, such a kernel does not have the form (\ref{tinv}) so that it would not be 
in ${\cal B}(E_{\infty})$.  Thus, the theorem provides new insights as to situations in which SDR methodologies will fail to capture the correct directions.

\noindent {\bf Remark 4.}  Each element of ${\cal B}(E_{\infty})$ will have a unique kernel associated with it and vice versa.   One example of a kernel that would exist in ${\cal B}(E_{\infty})$ is the Gaussian Kernel, whose kernel is given by 
$$K(z,\tilde z)= {\rm
exp}\{-\|z-\tilde z\ \| ^2/\rho\},$$
 where
$\|z- \tilde z\|^2 =\sum_{k=1}^p (z_k-\tilde z_k)^2$ and $\rho >0$ represents a scale parameter. The Gaussian
kernel generates the function space spanned by radial basis
functions, a complete overview for which can be found in B\"uhmann (2003).  Other examples of kernels that reside in ${\cal B}(E_{\infty})$ can be found in Table 1.

\begin{table}[htbp!]
\begin{center}
\caption{Examples of kernels that are members of ${\cal B}(E_{\infty})$.  Here $K_{\nu}$ denotes the modified Bessel
function of the second kind of order $\nu$.  }
\begin{tabular}{l l l}
\\
\hline
Kernel	&  K$(z,\tilde z)$	& Parameter ranges \\
\hline
Gaussian& $\exp\{-\|z-\tilde z\ \| ^2/\rho\}$ & $\rho > 0$\\
Mat\'ern& $\frac{2^{\nu-1}}{\Gamma(\nu)} \left ( \frac{\|z-\tilde z\|}{c} \right)^{\nu} K_{\nu}\left ( \frac{\|z-\tilde z\|}{c} \right) $& $c,\nu > 0$  \\
Generalized Cauchy &$\left [ 1 + \left ( \frac{\|z - \tilde z\|}{c} \right )^{\alpha} \right]^{-\tau/\alpha}$ & $c, \tau > 0,  0 < \alpha \leq 2$ \\
Dagum & \\
Powered Exponential & $\exp\{-\left  ( \frac{\|z-\tilde z\|}{c}\right)^{\alpha} \}$ & $c > 0, 0 < \alpha \leq 2$  \\
\hline
\end{tabular}
\end{center}
\end{table}

\section{Numerical Experiments}

An implication of the results from the previous section is that there will be qualitative similarity between results from linear and nonlinear sufficient dimension reduction methods.  To explore this, we performed the same simulation studies as in Lee et al. (2013).  
We consider two settings:
\begin{itemize}
    \item Setting 1 : $\sin(0.1\pi(X_1 + X_2)) + \epsilon$
    \item Setting 2 : $X_1/(1+e^{X_2}) + \epsilon$
\end{itemize}
where $\boldsymbol{X} = (X_1,\ldots,X_{p})$ $(p=10)$ are generated from a multivariate normal (MVN) distribution. Note that model (\ref{semiprop}) holds in Setting 1, but not in Setting 2.    We consider the following specifications for $\boldsymbol{X}$:
\begin{itemize}
    \item[(i)] $\boldsymbol{X} \sim V_1$, $V_1$ is MVN with mean 0 and covariance matrix $I_p$.
    \item[(ii)] $\boldsymbol{X} \sim V_2$, $V_2$ is a mixture distribution of MVN with mean $1_p = (1,\ldots,1)^T$ and covariance matrix $I_p$, and MVN with mean $-1_p = (-1,\ldots,-1)^T$ and covariance matrix $I_p$ with mixture probability 0.5 on each component. 
    \item[(iii)]  $\boldsymbol{X} \sim V_3$, $V_3$ is MVN with mean 0 and covariance matrix $0.6I_p + 0.41_p1_p^T$.
\end{itemize}
Hence (i) and (ii) represent scenarios with independent predictors while (iii) allows for dependence between predictors. Note that $\epsilon$ is independently generated from zero-mean normal distribution with variance $0.25$. We generate a training dataset with sample size 200, and we apply 4 sufficient dimension methods : sliced inverse regression (SIR) from Li (1991), Kernel sliced inverse regression (KSIR) from Wu (2008), KCCA (Kernel canonical correlation analysis) from Fukumizu et al. (2007) and GSIR (Generalized sliced inverse regression) from Lee et al. (2013). We then compute predictions on trained models to independently generated test data in each combination of setting with distribution of $\boldsymbol{X}$. To evaluate performance, we calculate two metrics, as in Lee et al. (2013): i) the absolute value of the Spearman correlation between the prediction from the trained models and true predictor in the test data, and ii) the absolute value of the Spearman correlation between the prediction from the trained models and response in the test data. Tables 2 and 3 show the mean and standard deviation of these Spearman correlations, averaged over simulations. \\
We used the Gaussian kernel for nonlinear sufficient dimension reduction. For estimation of the scale parameters in the Gaussian kernel, denoted by $\gamma_X$ and $\gamma_Y$, we compute
\begin{gather*}
    \hat{\gamma}_X = \frac{1}{2\hat{\sigma}_X^2} \quad \hat{\sigma}_X^2 = \binom{n}{2}^{-2} \sum_{i \ne j}||X_i - X_j||^2 \\
    \hat{\gamma}_Y = \frac{1}{2\hat{\sigma}_Y^2} \quad \hat{\sigma}_Y^2 = \binom{n}{2}^{-2} \sum_{i \ne j}|Y_i - Y_j|^2
\end{gather*}
For tuning parameters to calculate negative power of gram matrix, to prevent overfitting, we need to use a ridge regression-type parameter for $\boldsymbol{X}$ and response, say $\eta_X = \lambda_{max}(G_{\boldsymbol{X}})\zeta_X$ and $\eta_Y = \lambda_{max}(G_Y)\zeta_Y$. Here, $\lambda_{max}(A)$ is the maximum eigenvalue of matrix $A$ and $G_{\boldsymbol{X}}$, and $G_Y$ are the Gram matrices corresponding to the Gaussian kernels for $\boldsymbol{X}$ and $Y$, respectively, and   $\zeta_{\boldsymbol{X}}$ and $\zeta_{Y}$ take values between zero and one.    Alternatively, as suggested by Li (2018), they can be computed by generalized cross-validation (GCV), which minimizes errors $GCV_{\boldsymbol{X}}$ and $GCV_{Y}$  
\begin{gather*}
    GCV_{\boldsymbol{X}} = \frac{||G_Y - G_{\boldsymbol{X}}(G_{\boldsymbol{X}} + \zeta_{\boldsymbol{X}}\lambda_{max}(G_{\boldsymbol{X}})I_n)^{-1}G_Y||_F^2}{\{tr[I_n - G_{\boldsymbol{X}}(G_{\boldsymbol{X}} +\zeta_{\boldsymbol{X}}\lambda_{max}(G_{\boldsymbol{X}})I_n)^{-1} ]\}^2} \\
    GCV_{Y} = \frac{||G_{\boldsymbol{X}} - G_{Y}(G_{Y} + \zeta_{Y}\lambda_{max}(G_Y)I_n)^{-1}G_{\boldsymbol{X}}||_F^2}{\{tr[I_n - G_{Y}(G_{Y} +\zeta_{Y}\lambda_{max}(G_{Y})I_n)^{-1} ]\}^2}
\end{gather*}
where $||\cdot||_F^2$ denotes Frobenius norm. We run each simulation setting 400 times. We first use $\zeta_{\boldsymbol{X}} = \zeta_{Y} = 0.2$. Table 2 shows the absolute value of the Spearman correlation between the true predictors and response in this arbitrary setting of tuning parameters. We can see that performance between four methods are very similar in the independent predictors case. For dependent predictors, GSIR has the smallest correlation with true predictor. Next, we use grid points from (0.001,1) for both $\zeta_{\boldsymbol{X}}$ and $\zeta_Y$. Table 3 shows the absolute value of Spearman correlation between the true predictors and response in the controlled tuning parameters with GCV. In this setting, as before, the performance between four methods is similar although KCCA has the smallest correlation. Performance of GSIR is improved for the case of dependent predictors case compared to Table 2. \\
From these two simulation settings, we find our theory holds.  These results show that when we appropriately control tuning parameters, we obtain similar findings between linear and nonlinear sufficient dimension methods. 
\begin{table}[!ht]
    \centering
    \begin{tabular}{cccccccccc}
         &  & \multicolumn{4}{c}{Cor. with true predictor} &  \multicolumn{4}{c}{Cor. with response}\\ \hline
         & & SIR & KCCA & KSIR & GSIR & SIR & KCCA & KSIR & GSIR \\
    Setting 1 & (i) & 0.952 & 0.944 & 0.946 & 0.945 & 0.590 & 0.584 & 0.586 & 0.585 \\
    && (0.025) & (0.025) & (0.025) & (0.025) & (0.052) & (0.053) & (0.052) & (0.053) \\
    & (ii) & 0.972 & 0.966 & 0.969 & 0.965 & 0.71 & 0.706 & 0.708 & 0.705 \\
    && (0.014) & (0.016) & (0.014) & (0.017) & (0.038) & (0.039) & (0.038) & (0.04) \\
    & (iii) & 0.964 & 0.9 & 0.957 & 0.879 & 0.655 & 0.614 & 0.651 & 0.6 \\
    && (0.018) & (0.023) & (0.017) & (0.024) & (0.042) & (0.047) & (0.042) & (0.048) \\
    Setting 2 & (i) & 0.924 & 0.918 & 0.925 & 0.919 & 0.629 & 0.624 & 0.629 &0.624 \\
    && (0.021) & (0.024) & (0.021) & (0.024) & (0.048) & (0.049) & (0.048) & (0.049) \\ 
    & (ii) & 0.911 & 0.909 & 0.917 & 0.907 & 0.718 & 0.716 & 0.723 & 0.714 \\
    && (0.017) & (0.021) & (0.016) & (0.022) & (0.041) & (0.043) & (0.04) & (0.043) \\
    & (iii) & 0.926 & 0.803 & 0.909 & 0.764 & 0.632 & 0.6 & 0.625 & 0.535 \\
    && (0.022) & (0.045) & (0.031) & (0.046) & (0.046) & (0.055) &(0.048) & (0.057) \\
    \hline
    \end{tabular}
    \caption{Simulation results with setting tuning parameters $\zeta_{\boldsymbol{X}} = \zeta_Y = 0.2$}
    \label{tab:my_label}
\end{table}

\begin{table}[!ht]
    \centering
    \begin{tabular}{cccccccccc}
         &  & \multicolumn{4}{c}{Cor. with true predictor} &  \multicolumn{4}{c}{Cor. with response}\\ \hline
         & & SIR & KCCA & KSIR & GSIR & SIR & KCCA & KSIR & GSIR \\ 
         Setting 1 & (i) &  0.952 & 0.867 & 0.894 & 0.883 & 0.590 & 0.537 & 0.555 & 0.547 \\
         & & (0.025) & (0.05) & (0.038) & (0.046) & (0.052) & (0.061) & (0.057)& (0.059) \\
         & (ii) & 0.972 & 0.911 & 0.931 & 0.921 & 0.71 & 0.666 & 0.682 & 0.673 \\
         && (0.014) & (0.029) & (0.024) & (0.027) & (0.038) & (0.044) & (0.041) & (0.043) \\
         & (iii) & 0.964 & 0.948 & 0.947 & 0.953 & 0.655& 0.645 & 0.643 & 0.648 \\
         & & (0.018) & (0.023) & (0.022) & (0.020) & (0.042) & (0.045) & (0.045) & (0.044) \\
         Setting 2 & (i) & 0.924 & 0.856 & 0.888 & 0.871 & 0.629 & 0.59 & 0.61 & 0.6 \\
         && (0.021) & (0.053) & (0.03) & (0.048) & (0.048) & (0.059) & (0.052) & (0.056) \\ 
         & (ii) & 0.912 & 0.896& 0.913 & 0.905 & 0.718 & 0.722 & 0.731 & 0.728 \\
         && (0.017) & (0.032) & (0.023) & (0.029) & (0.041) & (0.043) & (0.041) & (0.042)\\
         & (iii) & 0.926 & 0.9 & 0.915 & 0.908 & 0.632 & 0.62 & 0.629 & 0.625 \\
         && (0.022) & (0.046) & (0.028) & (0.045) & (0.046) & (0.051) & (0.048) & (0.051) \\ \hline
    \end{tabular}
    \caption{Simulation results with GCV for tuning parameters $\zeta_{\boldsymbol{X}}$ and $\zeta_Y$}
    \label{tab:my_label}
\end{table}

\section{Discussion}

In this note, we have shown a new asymptotic framework using nested function spaces applied to sufficient dimension reduction.   There are several implications of our work.  First, it shows that a new and separate justification for the types of kernels considered and used in Li et al. (2011) and Lee et al. (2013) is that they represent approximators to the space of distributions that satisfy the linearity condition.  We show explicitly that the linearity condition induces a function space in a certain limiting sense that requires radially symmetric kernels (i.e., kernels that depend on the absolute magnitude of interpoint distances).    Conversely, kernel machines that are not radially symmetric do not live in $B(E_{\infty})$.  

There are several future directions currently under investigation.  First, we would like to better understand the error in approximating model (\ref{semiprop}) by a function in $B(E_{\infty})$.  Second, we would like to understand how this work can be extended to handle a mixture of discrete and continuous predictors.  As is well-studied in the sufficient dimension reduction literature, directions are not estimable with discrete covariates.  While Chiaromonte et al. (2002) developed an approach for sliced inversed regression that accommodates categorical covariates, extending it to the nonlinear setting, such as what is described in the current setting, remains an open problem. 

\newpage

\begin{flushleft}
{\bf References}
\end{flushleft}
\begin{description}

\item{} Aronszajn, N. 1950. Theory of reproducing kernels. { Transactions of the American Mathematical Society} {68}, 337 -- 404. 
%doi:10.1090/S0002-9947-1950-0051437-7. JSTOR 1990404. MR 51437.

\item{} Berlinet, A., Thomas-Agnan, C. 2004.  {Reproducing kernel Hilbert spaces in Probability and Statistics.} New York: Kluwer Academic Publishers

\item{} Brillinger, D. 1977.  The identification of a particular nonlinear time series system. Biometrika 64 509–515.

\item{} B\"uhmann, M. D. 2003. {Radial Basis Functions: Theory and Implemetations}.   Cambridge: Cambridge University Press.

\item{} Cook, R. D. 1998.  Regression Graphics.  Wiley, New York.

\item{} Cook, R. D.,  Ni, L. 2005. Sufficient dimension reduction via inverse regression: A minimum discrepancy approach.  Journal of the American Statistical Association 100, 410 -- 428.

%\item{} Cook, R. D. 2007.  Fisher Lecture: dimension reduction in regression.  {\it Statist. Sci.} {\bf 22}, 1 -- 26.
%\item{} Cook, R. D., Li, B. and Chiaromonte, F. 2007.  Dimension reduction in regression without matrix inversion.  {\it Biometrika} {\bf }, 569 -- 584

%\item Cristianini, N.,  Shawe-Taylor, J. 2000.  {\it An
%Introduction to Support Vector Machines and Other Kernel-based
%Learning Methods.}  Cambridge: Cambridge University Press.

% \item{} Freund, Y. and Schapire, R. E. 1997.  A decision-theoretic generalization of on-line learning and an application to boosting.  {\it Journal of Computer and System Sciences} {\bf 55}, 119--139.
 
%\item{} Hastie, T., Tibshirani, R. and Friedman, J. 2009.  The Elements of Statistical Learning: Data Mining, Inference, and Prediction. Second Edition.  New York: Springer.  
 
%\item{}{Helland, I. S. 1988.  On the structure of partial least squares regression.  {\it Communications in Statistics -- Simulation and Computation} {\bf 17}, 581 -- 607.}

\item{} Ferr\'e, L., Yao, A. F. 2003. Functional sliced inverse regression analysis. Statistics 37, 475 -- 488.

\item{} Fukumizu, K., Bach, F. R.,  Jordan, M. I. 2004. Dimensionality reduction for supervised learning with reproducing kernel Hilbert spaces.  Journal of Machine Learning Research 5, 73 -- 99.

\item{} Fukumizu, K., Bach, F. R., Gretton, A. 2007. Statistical consistency of kernel canonical correlation analysis. Journal of Machine Learning Research, 8, 361 --383.

\item{} Fukumizu, K., Bach, F. R.,  Jordan, M. I. 2009. Kernel dimension reduction in regression.  The Annals of Statistics 37,  1871 -- 1905. 

%\item{} Gneiting, T. 2013.  Strictly and non-strictly positive definite functions on spheres.   { Bernoulli} {19}, 1327 -- 1349.
 
\item{} Lee, K-Y., Li, B., Chiaromonte, F.  2013.  A general theory for nonlinear sufficient dimension reduction: formulation and estimation. The Annals of Statistics 41  221--249. 

\item{} Li, B., Artemiou, A., Li, L. 2011. Principal support vector machines for linear and nonlinear sufficient dimension reduction. The Annals of Statisics 39, 3182 -- 3210.

\item{} Li, B. 2018. Sufficient dimension reduction: Methods and applications with R. CRC Press.

\item{} Li, K. C. 1991.  Sliced inverse regression for dimension
reduction (with discussion).  {Journal of the American
Statistical Association} {86}, 316 -- 342.

\item{} Li, K. C. 1992.  On Principal Hessian Directions for data visualization and dimension reduction: another application of Stein's lemma.  {Journal of the American
Statistical Association} {87}, 1025 -- 1039.

\item{} Li, K. D., Duan, N. 1989.  Regression analysis under link violation.  The Annals of Statistics 17, 1009 -- 1052.

%\item{} Ma, Y., Zhu, L. 2012.  A semiparametric approach to dimension reduction.   { Journal of the American Statistical Association} {107},  168-179.

%\item{} Robins, J. 1995.  Discussion of `Causal diagrams in
%empirical research' by J. Pearl.  {\it Biometrika} {\bf 82}, 695
%-- 698.

\item{} Schoenberg, I. J. 1938.  Metric spaces and completely monotone functions.  {Annals of Mathematics} {39}, 811 -- 841.

%\item{} Xia, Y., Tong, H., Li, W. K. and Zhu, L. X. 2002 An adaptive estimation of dimension
%reduction space with discussion. Journal of the Royal Statistical Society, Series B, 64,
%363-410.

\item{} Wu, H. M. 2008. Kernel sliced inverse regression with applications
to classification. Journal of Computational and Graphical Statistics 17, 590-610. 

\item{} Wu, Q., Liang, F., Mukherjee, S. Kernel sliced inverse regression: regularization and consistency. 2013. Abstract and Applied Analysis. doi:10.1155/2013/540725. %https://projecteuclid.org/euclid.aaa/1393450311

\item{} Yin, X., Li, B., Cook, R.D.  2008. Successive dimension extraction for estimating the central subspace in multiple-index regression. Journal of Multivariate Analysis 99, 1733--1757.

\end{description}

\end{document}